# Negative Differential Conductivity in Carbon Nanotubes in the Presence of an External Electric Field

S. S. Abukari[a], S.Y.Mensah[a] N. G. Mensah[b, *], K.W. Adu[c,d] K. A. Dompreh[a], and A. K. Twum[a]

[a]*Department of Physics, Laser and Fibre Optics Centre, University of Cape Coast, Cape Coast, Ghana*

[b]*Department of Mathematics, University of Cape Coast, Cape Coast, Ghana*

[c]*Department of Physics, Penn State Altoona, Altoona, Pennsylvania 16601, U.S.A*

[d]*Material Research Institute, The Pennsylvania State University, University Park 16802, U.S.A*

[*]*Corresponding author.* S.Y.Mensah

Tel.:+233 042 33837

*E-mail address*: profsymensah@yahoo.co.uk

**Abstract**

We study theoretically the electron transport properties in carbon nanotubes under the influence of an external electric field $E(t)$ using Boltzmann's equation. The current-density equation is derived. Negative differential conductivity is predicted when $\omega\tau \ll 1$ (quasi-static case). We observed this in the neighbourhood where the constant electric field $E_o$ is equal to the amplitude of the AC electric field $E_1$ and the peak decreases with increasing $E_1$. This phenomenon can also be used for the generation of terahertz radiation without electric current instability.

## 1. Introduction

Carbon nanotubes (CNs) were first discovered in 1991 [1], and since then great deal of interest has been focused on these quasi-one-dimensional monomolecular structure because of their unique electrical, mechanical, and chemical properties. Nonlinear effects in CNs are of great interest for potential applications in nanoelctronics.

Negative differential conductivity (NDC) has been predicted in CNs at room temperature under the condition, when $\kappa_B T > \varepsilon_c, \Delta\varepsilon$ in a certain range of electric field strength [2]. The NDC is believed to provide current instability in CNs which is destructive for the formation of terahertz (THz) radiation as in semiconducting superlattices. Simultaneously applied both dc-and ac-fields will result in nonlinear phase of the instability as is observed in semiconducting superlattices (SL). Mensah [3] studied the negative differential effect in a semiconductor SL in the presence of an external electric field. The theory indicated that the current-density electric field characteristic shows a negative differential conductivity when $\omega\tau \ll 1$ and this occurs in the neighbourhood where the constant electric field $E_o$ is equal to the amplitude of the ac electric field $E_1$ and the peak decreases with increasing $E_1$. The theory agrees fairly well with an experiment [4] that indicated "right shift" of the IV maximum, which is typical for a SL without domain formation. Reference [4] demonstrated ultrafast creation and annihilation of space-charge domains in a semiconductor superlattice observed by use of Terahertz fields.

Up to now, NDC has been observed only in a d.c electric field in both doped and undoped CNs. We shall, in this paper, show that this possible in a d.c and a.c electric fields.

This work will be organised as follows: section 1 deals with introduction; in section 2, we establish the theory and solution of the problem; section 3, we discussion the results and draw conclusion.

2. **Theory**

We consider a response of electrons in an undoped single-wall achiral CNS (ie zigzag or armchair ) to the action of a strong pump field.

$$E(t) = E_o + E_1 cos\omega t \tag{1}$$

Where the dc bias $E_o$ is small and the ac field is quasistatic, $\omega\tau \ll 1$.

The investigation is done within the semiclassical approximation in which the motion of $\pi-$electrons are considered as classical motion of free quasi-particles in the field of crystalline lattice with dispersion law extracted from quantum theory. Taking into account the hexagonal crystalline structure of a rolled grapheme in a form of CNs and using the tight binding approximation, the energy dispersion is expressed as

$$\varepsilon(s\Delta p_\varphi, p_z) \equiv \varepsilon_s(p_z) = \pm\gamma_0 \left[1 + 4cos(ap_z)cos\left(\frac{a}{\sqrt{3}}s\Delta p_\varphi\right) + 4cos^2\left(\frac{a}{\sqrt{3}}s\Delta p_\varphi\right)\right]^{1/2} \tag{2}$$

for the zigzag CNs and

$$\varepsilon(s\Delta p_\varphi, p_z) \equiv \varepsilon_s(p_z) = \pm\gamma_0 \left[1 + 4cos(ap_z)cos\left(\frac{a}{\sqrt{3}}s\Delta p_\varphi\right) + 4cos^2\left(\frac{a}{\sqrt{3}}s\Delta p_\varphi\right)\right]^{1/2} \tag{3}$$

for the armchair CNs [2].

Where $\gamma_0 \sim 3.0eV$ is the overlapping integral, $p_z$ is the axial component of quasimomentum, $\Delta p_\varphi$ is transverse quasimomentum level spacing and $s$ is an integer. The expression for $a$ in Eqs. (2) and (3) is given as $a = 3b/2\hbar$, $b = 0.142nm$ is the C-C bond length. The – and + signs correspond to the valence and conduction bands respectively. Due to the transverse quantization of the quasi-momentum, its transverse component can take $n$ discrete values, $p_\varphi = s\Delta p_\varphi = \pi\sqrt{3}\ \pi/an\ (s = 1\dots, n)$. Unlike transverse quasimomentum $p_\varphi$, the axial quasimomentum $p_z$ is assumed to vary continuously within the range $0 \leq p_z \leq 2\pi/a$ , which corresponds to the model of infinitely long CN($L = \infty$). This model is applicable to the case under consideration because of the restriction to the temperatures and /or voltages well above the level spacing [5], ie. $k_B T > \varepsilon_C$, $\Delta\varepsilon$ , where $k_B$ is Boltzmann constant, $T$ is the temperature, $\varepsilon_C$ is the charging energy. The energy level spacing $\Delta\varepsilon$ is given by

$$\Delta\varepsilon = \pi\hbar v_F/L \tag{4}$$

where $v_F$ is the Fermi velocity and $L$ is the carbon nanotube length [6]

Employing Boltzmann equation with a single relaxation time approximation

$$\frac{\partial f(p)}{\partial t} + eE(t)\frac{\partial f(p)}{\partial P} = -\frac{[f(p) - f_0(p)]}{\tau} \tag{5}$$

where $e$ is the electron charge, $f_0(p)$ is the equilibrium distribution function , $f(p,t)$ is the distribution function, and $\tau$ is the relaxation time. The electric field $E$ is applied along CNs axis. In this problem the relaxation term $\tau$ is assumed to be constant. The justification for $\tau$ being constant can be found in [7]. The relaxation term of Eq. (5) describes the effects of the dominant type of scattering (e.g. electron-phonon and electron-twistons) [8]. For the electron scattering by twistons (thermally activated twist deformations of the tube lattice), $\tau$ is proportional to $m$ and the $I - V$ characteristics have shown that scattering by twistons increases $E^{max}$ and decreases $|\partial j_z/\partial E_z|$ in the NDC region; the lesser $m$, the stronger this effect. Quantitative changes of the $I - V$ curves turn out to be insignificant in comparison with the case of $\tau = $ const [7, 8].

Expanding the distribution functions of interest in Fourier series as;

$$f_0(p) = \Delta p_\varphi \sum_{s=1}^{n} \delta(p_\varphi - s\Delta p_\varphi) \sum_{r\neq 0} f_{rs}\ e^{iarp_z} \tag{6}$$

and

$$f(p,t) = \Delta p_\varphi \sum_{s=1}^{n} \delta(p_\varphi - s\Delta p_\varphi) \sum_{r\neq 0} f_{rs}\ e^{iarp_z} \emptyset_\upsilon(t) \tag{7}$$

Where the coefficient, $\delta(x)$ is the Dirac delta function, $f_{rs}$ is the coefficient of the Fourier series and $\emptyset_\upsilon(t)$ is the factor by which the Fourier transform of the nonequilibrium distribution function differs from its equilibrium distribution counterpart.

$$f_{rs} = \frac{a}{2\pi\Delta p_\varphi S} \int_0^{\frac{2\pi}{a}} \frac{e^{-iarp_z}}{1 + exp(\varepsilon_s(p_z)/k_B T)} dp_z \qquad (8)$$

Substituting Eqs. (6) and (7) into Eq. (5), and solving with Eq. (1) we obtain

$$\emptyset_\upsilon(t) = \sum_{k=-\infty}^{\infty} \sum_{m=-\infty}^{\infty} \frac{J_k(r\beta)J_{k-\upsilon}(r\beta)}{1 + i(earE_0 + k\omega)\tau} exp(i\upsilon\omega t) \qquad (9)$$

where $\beta = \frac{eaE}{\omega_1}$, $J_k(\beta)$ is the Bessel function of the $k^{th}$ order and $\Omega = eaE_0$.

Similarly, expanding $\varepsilon_s(p_z)/\gamma_0$ in Fourier series with coefficients $\varepsilon_{rs}$

$$\frac{\varepsilon_s(p_s, s\Delta p_\varphi)}{\gamma_0} = \varepsilon_s(p_z) = \sum_{r \neq 0} \varepsilon_{rs} e^{iearp_z} \qquad (10)$$

Where

$$\varepsilon_{rs} = \frac{a}{2\pi\gamma_0} \int_0^{\frac{2\pi}{a}} \varepsilon_s(p_z) e^{-iearp_z} dp_z \qquad (11)$$

and expressing the velocity as

$$\upsilon_z(p_z, s\Delta p_\varphi) = \frac{\partial \varepsilon_s(p_z)}{\partial p_z} = \gamma_0 \sum_{r \neq 0} iar\, \varepsilon_{rs} e^{iearp_z} \qquad (12)$$

We determine the surface current density as

$$j_z = \frac{2e}{(2\pi\hbar)^2} \iint f(p)\, \upsilon_z(p) d^2 p,$$

or

$$j_z = \frac{2e}{(2\pi\hbar)^2} \sum_{s=1}^{n} \int_0^{\frac{2\pi}{a}} f\left(p_z, s\Delta p_\varphi, \emptyset_\upsilon(t)\right) \upsilon_z(p_z, s\Delta p_\varphi) dp_z \qquad (13)$$

and the integration is taken over the first Brillouin zone. Substituting Eqs. (7), (9) and (12) into (13) we find the current density for the zigzag CNs after averaging over a period of time $t$, we obtain

$$j_z = \frac{8e\gamma_0}{\sqrt{3}\hbar n a_{c-c}} \sum_{r=1}^{\infty} r \sum_{k=-\infty}^{\infty} \frac{J_k^2(r\beta)(\Omega r + k\omega)\tau}{1 + \left((\Omega r + k\omega)\tau\right)^2} \sum_{s=1}^{n} f_{rs}\varepsilon_{rs} \qquad (14)$$

For $\omega\tau \leq 1$, Eq. (14) can be re-written in the form of Ref. [3] as;

$$j_z = \frac{8e\gamma_0}{\sqrt{3}\hbar n a_{c-c}} \sum_{r=1}^{\infty} r \left( \frac{\frac{1}{4}\{[1 + (Z_c + \beta)^2]^{1/2} + [1 + (Z_c - \beta)^2]^{1/2}\}^2 - \beta^2 - 1}{[1 + (Z_c + \beta)^2][[1 + (Z_c - \beta)^2]]} \right)^{1/2}$$

$$\times \sum_{s=1}^{n} f_{rs}\varepsilon_{rs} \qquad (15)$$

$Z_c = earE_o$ and $\beta = \frac{earE_1}{\omega}$.

### 3. Results, Discussion and Conclusion

The current density expression in zigzag CNs subjected to dc bias field $E_o$ and quasistatic ac field ($\omega\tau \ll 1$) is obtained by using the solution of the Boltzmann equation with constant relaxation time $\tau$.

We observed that the current density $j_z$ is a function of the electric field $E_o$ and $E_1$. We illustrated how these parameters affect $j_z$ using Matlab. Fig. 1 represents the graph of $j_z/j_o$ on $E_o$ for $\beta = 2, 4, and\ 8$ at $\omega\tau = 0.2$. Fig. 1(a) represents the armchair CNs and (b) superlattices.. The figures show the linear dependence of $j_z$ on $E_o$ at weak strengths of the electric of the external field (i.e. the region of ohmic conductivity). As $E_o$ increases, the current density $j_z/j_o$ increases and at $E_o = E_o^{max}$ the current density reaches a maximum value $(j_z/j_o)^{max}$. Further increase of $E_o$ results in the decrease of the $j_z/j_o$> Thus, the region of negative differential conductivity (NDC) where $\partial j_z/\partial E_o < 0$. We noted that in the case investigated there is a shift of the maximum of the current density electric field curves towards larger $E_o$ values. This "right shift" is caused by a nonliearity of the Esaki-Tsu characteristics which is very strong in CNs because of the high stark component (summation over $r$). The role of the high stark components in CNs is essential and intergral nonlinearity of the CNs is much higher than in SL [9, 10].The shift increases with increasing the amplitude of the ac field.

The $I-V$ curves are qualitatively similar for the CNs and the superlattices (see Fig.1). However, the NDC effect in SL appeared at larger field strengths comparing with the CNs.

The estimations of the restrictions of the theoretical approach used can be found in [11]. From expression (15) a graph of $j_z/j_o - E_o$ is plotted and it is observed that $j_z$ assumes its maximum value in the vicinity of $Z_c \cong \beta$ for any given value of $\beta$. This indicates that NDC is observed where the constant field $E_o$ is approximately equal to the amplitude of the AC electric field $E_1$. It is quite interesting to note that the graphs of expression (14) and (15) are qualitatively the same for $\omega\tau \leq 1$. See Fig. 2 and the peaks of the curves decreases with increasing $E_1$.

In conclusion we have studied theoretically the current-density electric field characteristic in the presence of ac-dc driven field and negative differential conductivity was observed. The current-density electric field characteristic shows a negative differential conductivity when ($\omega\tau \ll 1$) (quasi-static case). This occurs in the neighbourhood where the constant electric field $E_o$ is equal to the amplitude of the AC electric field $E_1$ and the peak decreases with increasing $E_1$. We suggest that this phenomenon can also be used for the generation of terahertz radiation without electric current instability.

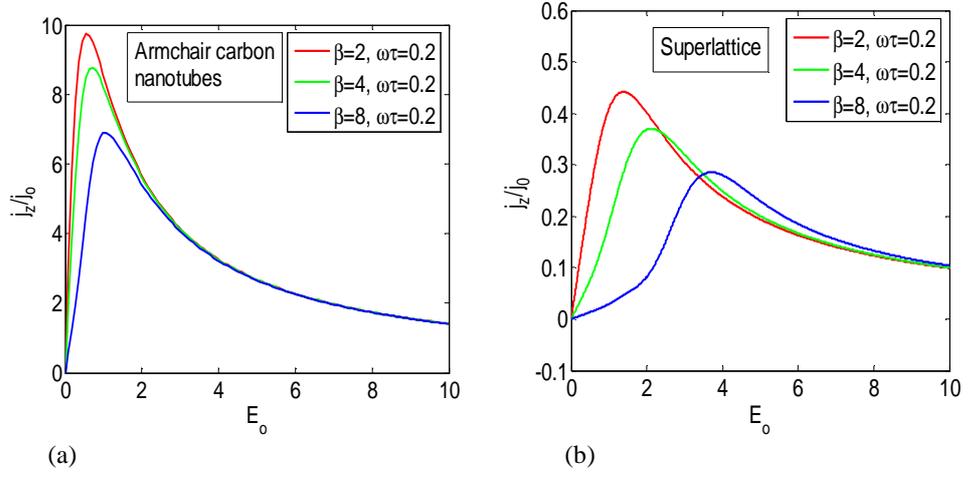

**Fig. 1:** $j_z/j_o - E_o$ curves for (a) armchair and (b) superlattice when: (—) $\omega\tau = 0.2, a = 2$ ; (—) $\omega\tau = 0.2, a = 4$; (—) $\omega\tau = 0.2, a = 8$.

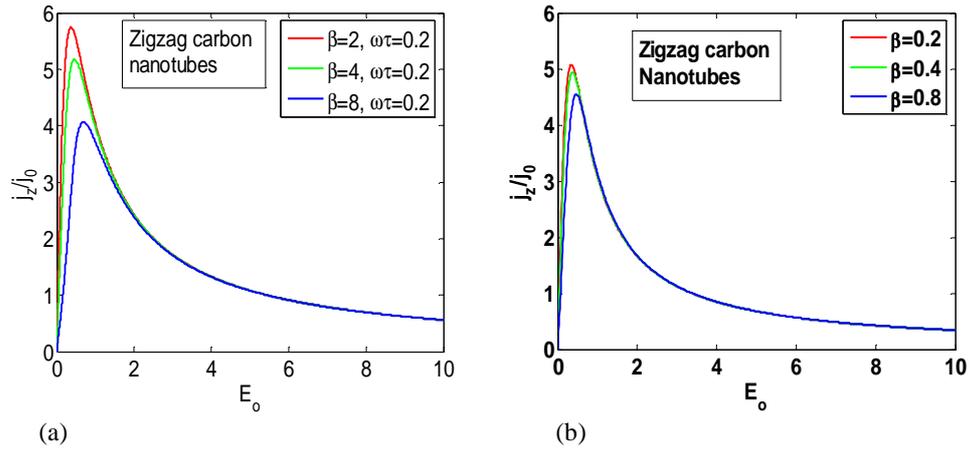

**Fig. 2:** $j_z/j_o - E_o$ curves for (a) Expression (14); when: (—) $\omega\tau = 0.2, a = 2$ ; (—) $\omega\tau = 0.2, a = 4$; (—) $\omega\tau = 0.2, a = 8$ and (b) Expression (15); when: (—) $\beta = 0.2$; (—) $\beta = 0.4$; and (—) $\beta = 0.8$

**Reference**

[1] S. Iijima, Nature (London) 354, 56 (1991).
[2] A. S. Maksimenko and G. Ya. Slepyan, Phys.Rev. Lett. 84 362 (2000)
[3] S.Y. Mensah, Journal Physics Condens Matter 4 (1992) L325-329.
[4] F. Klappenberger, K. N. Alekseev, K.F. Renk, R. Scheurer, E. Schomburg, S. J. Allen, G. R. Ramian, J. S. S. Scott, A. Kovsh, V. Ustinov, and A. Zhokov, Eur. Phys. J. B 39, 483-489,2004.
[5] C. Kane, L. Balents, and M. P. A. Fisher, Phys. Rev. Lett. **79**, 5086 (1997).
[6] M. F. Lin, and K. W.K. Shung, Phys. Rev. B 52, pp. 8423–8438, 1995.
[7] C. L Kane, E. J. Mele, R. S. Lee, J. E. Fischer, P. Petit, H. Dai,. A. Thess, R. Smalley, E. A. R. M. Verscheueren, S. J. Tans, and C. Dekker, Europhys. Lett. 41, 683-688 (1998).
[8] R. A. Jishi, M. S. Dresselhaus, and G. Dresselhaus, Phys. Rev. B 48, 11385 - 11389, (1993).
[9] F. G. Bass and A. A. Bulgakov, *Kinetic and Electrodynamic Phenomena in Classical and Quantum Semiconductor Superlattices* (Nova, New York,1997)
[10] O. M. Yevtushenko *et al* Phys. Rev. Lett 79, 1102 (1997); G. Ya. Slepyan *et al*. Phys. Rev. B **57**, 9485 (1998)
[11] G. Ya. Slepyan, S. A. Maksimenko, V. P. Kalosha, J. Herrmann, E. E. B. Campbell, I. V. Hertel, Phys. Rev. A **60**, 2 (1999).